\documentclass[11pt]{article}

\input epsf.sty

\usepackage{graphicx,amsmath,amssymb}
\def\lromn#1{\uppercase\expandafter{\romannumeral#1}}

\begin{document}

\begin{flushright}
\today \\
\end{flushright}

\begin{center}
\begin{Large}

{\bf Neutrino pair and gamma beams from circulating excited ions}
\end{Large}

\vspace{2cm}
M.~Yoshimura and N. Sasao$^{\dagger}$

\vspace{0.5cm}
Center of Quantum Universe, Faculty of
Science, Okayama University \\
Tsushima-naka 3-1-1 Kita-ku Okayama
700-8530 Japan

\vspace{0.2cm}
$^{\dagger}$
Research Core for Extreme Quantum World,
Okayama University \\
Tsushima-naka 3-1-1 Kita-ku Okayama
700-8530 Japan \\

\end{center}

\vspace{5cm}

\begin{center}
\begin{Large}
{\bf ABSTRACT}
\end{Large}
\end{center}

We propose a new method of producing neutrino pair beam that consists of
a mixture of neutrinos and anti-neutrinos of all flavors.
The idea is based on a  coherent neutrino pair emission
from excited ions in circular motion.
High energy gamma ray much beyond the keV range may also be produced
by a different choice of excited level.

\vspace{4cm}
%\pacs{ 
PACS numbers
\hspace{0.5cm} 
13.15.+g, %% Neutrino interactions  
14.60.Pq, %% Neutrino mass and mixing 

Keywords
\hspace{0.5cm} 
neutrino  beam, neutrino pair emission, excited heavy ion, 
coherent gamma ray beam

\newpage

\section
{\bf Introduction}

Synchrotron radiation is a very useful tool
of  photon emission up to the X-ray energy range, providing a well collimated beam.
We examine a similar problem of neutrino pair emission under
a circular motion of  ions.
When excited ions with a high coherence are
circulated, emission rates become large with neutrino
energies extending  to much beyond the keV region in the form of well collimated beam.
Produced neutrino beam is a  mixture of
all pairs of neutrinos, including $\nu_{\mu} \bar{\nu}_{\mu}\,, \nu_{\tau} \bar{\nu}_{\tau}$.
This gives a CP-even neutrino beam, hopefully
providing an ideal setting to
test fundamental symmetries of particle physics \cite{cpt theorem and sterile},
in particular, to
 measure CP violating (CPV) phases in the neutrino sector \cite{cp violation},
\cite{cpt theorem and sterile}, \cite{barger et al}, \cite{pakvasa}.
Circulation of highly stripped heavy ions is desirable to achieve the highest
neutrino energy in the GeV region with the largest production rates.

Our method of calculation may be adapted to synchrotron radiation 
that occurs at electron machine,
giving essentially the same results as in \cite{schwinger},
although our method of calculation is different.
We shall make it clear how a GeV range intense beam of
neutrino pairs is made possible if one uses excited
ions instead of ions in the ground state.

One may also produce high energy gamma ray much beyond the keV range
by an appropriate choice of excited level of different parity.
This may be very useful since the usual electron synchrotron
can only produce the keV range photon.

The rest of this paper is organized as follows.
In the first two sections we shall explain our semi-classical approximation 
to treat the ionic motion as given classically and to calculate
the probability and its rate of neutrino pair emission in the standard electroweak theory.
In Section 4 we give the core  calculation of a phase integral that
appears in the rate calculation.
We find that with excited ions the phase integral over time contains
stationary points of the phase, leading to large neutrino pair emission rates.
In the following section we compute the differential  energy spectrum
of neutrino pair production at synchrotron site.
In Section 6 we discuss 
a similar problem of photon emission
from electric dipole allowed atomic transition.
When a good coherence among ions in the excited and the ground levels
is prepared and maintained, it might even be possible to have a coherent
gamma ray emission much like laser in the optical region.

In a sequel paper we shall discuss neutrino oscillation 
experiments that can be done away from the synchrotron.

Throughout this work we use the natural unit of $\hbar = c = 1$.

\vspace{0.5cm} 
\section
{\bf Semi-classical approximation}

The total wave function of a composite ion
consists of a direct product of the central motion (CM) part
of ion as a whole and its internal part
as a consequence of separation of hamiltonian operator into an independent sum
of two terms.
For the neutrino pair emission process of internal atomic transition,
$|e \rangle \rightarrow | g \rangle$,
 the other CM hamiltonian part never contributes simultaneously.
It only contributes to  the cases of $|a\rangle \rightarrow |a \rangle \,, a=e,g$,
and this gives rise to the usual synchrotron emission
in much the same way as in the electron machine.
For the internal transition,
the CM part of wave function $\Psi_i$
gives a weight factor of its probability density $|\Psi_i|^2 = 1/(\gamma V)$
($V$ the quantization volume) in the internal part of hamiltonian, 
in accordance to the general rule of the correct property of
the lifetime under the
Lorentz transformation $\propto \gamma$ \cite{nishijima}.
Here $\gamma$ is the boost factor of excited ion related to the constant velocity $v$
of circular motion 
by $v=\sqrt{1-1/\gamma^2} \sim 1 - 1/(2\gamma^2)$.
Strictly, one needs the instantaneous boost factor $\gamma(t)$
of a time dependent function,
but the emission region around the circular orbit is short, and one may replace this
by the constant circular velocity.
For the ion internal state
we shall confine ourselves to a two-level system
 as an approximation, its ionic states being
 $|e \rangle $ and $ |g\rangle$.
The metastable state $|e \rangle$ in an upper energy level
is assumed to have the same parity as that of the ground state $|g \rangle$
such that fast electric dipole transition is forbidden, while a
magnetic dipole (M1) transition and the neutrino pair emission are both allowed.
Relevance of the M1 transition to neutrino pair emission is explained in due course.
Another electric dipole case between different parity states is useful for high energy gamma ray
emission and is discussed in Section 6.

Bilinear forms of wave functions such as the  ion current 
may be described  in terms of the density matrix,
$\rho_{ab}\,, a,b = e, g$  for its internal part.
We assume that the central motion is described by
the classical trajectory function $x_A(t)$
($t$  the time at observation) of circular motion.
The density matrix for two-level system is governed by
the optical Bloch equation.
Its solution may readily be derived in terms of initial values.
In particular, the off-diagonal 
element $\rho_{eg}(t)$ of our system may be described to a good approximation
\cite{two-level} by
\begin{eqnarray}
&&
\rho_{eg}(t) = \rho_{eg}(0) \exp[- (i \epsilon_{eg} + \frac{1}{T_2}) \frac{t}{\gamma}]
\,,
\label {phase for de-excitation}
\end{eqnarray}
when effects of photon emission are highly suppressed.
Throughout this work we use the time in the laboratory system
in which measurements of neutrino beam experiments are done.
The phase relaxation rate $1/T_2$ is usually  larger than
its minimum value $1/(2\tau_e) $ ($\tau_e$ being
the natural lifetime of state $|e\rangle$) 
that occurs when the phase relaxation is dominated by the spontaneous
decay.

In the atomic physics community the quantity $\rho_{eg}$
is called the coherence.
For a pure quantum state of a single atom, it is given by a
quantum mixture of two states, $| e\rangle$ and $| g \rangle$.
Its value is  bounded 
to be less than the value $1/2$ in our normalization convention.
Its macroscopic average over a collective body of atoms or ions
is usually much less than this maximum value.
We shall not discuss the experimental problem
of how a large initial coherence given by $\rho_{eg}(0)$ may be prepared.

We can neglect contributions of ionic states that remain   in either the
excited or the ground state, their rates being proportional to $\rho_{ee}^2\,, \rho_{gg}^2$,
since they give rise to neutrino pair emission of much smaller rates and
much smaller energies, the neutrino-pair analogue of the usual synchrotron radiation.
This result originates from that these density matrix elements
have no oscillating phase factor as in $\rho_{eg}(t)\propto e^{- i \epsilon_{eg}t/\gamma} $.
It is found that both in this case and in the case of electron synchrotron
no large neutrino pair production occurs, as is made evident below.

\vspace{0.5cm} 
\section
{\bf Perturbation theory of neutrino pair emission}

In our semi-classical approximation
the hamiltonian system of interacting neutrino with atomic electrons  is quadratic in neutrino field variables,
and one can readily solve the problem of neutrino pair emission, using the perturbation theory of the weak coupling $G_F$.
The four-Fermi interaction of neutrinos and atomic electrons is given by
the hamiltonian (written in terms of neutrino field operators):
\begin{eqnarray}
&&
H_w^{(0)} = \int d^3x \frac{G_F}{\sqrt{2}}  \cdot \sum_{i,j = 1,2,3} 
( c_{ij}^V V^{\beta}(x) +  c_{ij}^A A^{\beta}(x)\,)
\nu_i^{\dagger}(x) \sigma_{\beta} \nu_j(x)
\,,
\label {fermi interaction}
\end{eqnarray}
with $(\sigma_{\beta}) = (1, - \vec{\sigma})$.
We use the neutrino index convention of Roman alphabets, $a,b,c$,
to indicate neutrino flavor states, $\nu_e, \nu_{\mu}, \nu_{\tau}$,
and  Roman alphabets, $i, j, k$, to indicate
mass eigenstates $\nu_1, \nu_2, \nu_3$.
The neutrino mass ordering is taken as usual:
$m_3 > m_2 > m_1$ for the normal hierarchy case and
$ m_2 > m_1 > m_3$ for the inverted hierarchy case.
Both W- and Z-boson exchange contributions are added,
and the hamiltonian is written in the Fierz-transformed form (charge retention ordered).
There are both vector and axial-vector currents, $V(x), A(x)$, with their
couplings $c_{ij}^{V,A}$.
We may  assume the non-relativistic limit for transitions of internal electron states in
the rest frame of ion,
which singles out as the dominant contribution the spatial part of axial 4-vector 
$c_{ij}^A S^{\beta}$
in the form of the electron spin current:
\(\:
 (0\,,  \vec{S}_e ) \,,
\vec{S}_e = \sum_a \langle g |\vec{\sigma}_a/2| e \rangle
\,,
\: \)
where the sum  is taken over valence electrons of ions.
Note that the monopole term of the vector part $\propto c_{ij}^V V_0$ vanishes
due to the orthogonality of wave functions between $|e\rangle $ and $|g \rangle$.
The coefficients of  axial-vector parts are 
\begin{eqnarray}
&&
C \equiv ( c_{ij}^A) 
\,, \hspace{0.5cm}
 c_{ij}^A= U_{ei} U^*_{ej} - \frac{1}{2}\delta_{ij}
\,, \hspace{0.5cm}
CC^{\dagger} = \frac{1}{4}
\,,
\end{eqnarray}
in the standard electroweak theory.

In the laboratory frame the relevant current becomes  \cite{jackson}
\begin{eqnarray}
&&
(S_{\alpha}) = (\gamma \vec{\beta}\cdot \vec{S}_e\,, \vec{S}_e +
\frac{\gamma^2}{\gamma+1} (\vec{\beta}\cdot\vec{S}_e) \vec{\beta}\, )
\sim \gamma ( \vec{\beta}\cdot \vec{S}_e\,, (\vec{\beta}\cdot\vec{S}_e) \vec{\beta}\,)
\,,
\label {helcity summed spin factor}
\end{eqnarray}
where $\vec{\beta}$ is the Lorentz boost vector.
Averaging over atomic spin direction to the leading order of large $\gamma$ gives the squared amplitude 
summed over neutrino helicities \cite{my-prd 07},
\begin{eqnarray}
&&
\gamma^2 \frac{S_e^2}{3}
\left( 1+ \frac{1}{3} \frac{\vec{p}_1 \cdot \vec{p}_2}{E_1 E_2} - \frac{m_1m_2}{2E_1 E_2} \delta_M
\right)
\,,
\end{eqnarray}
where $\delta_M = 1$ for the Majorana neutrino and $\delta_M = 0$ for the Dirac neutrino.
Our experience of calculations for heavy atoms such as Xe, Yb, etc
\cite{renp overview}, \cite{renp pv} shows that these matrix elements $\vec{S}_{e}$
are of order unity or O(0.1) where the intermediate coupling
scheme of heavy atoms holds.
We assume in the following that the M1 transition matrix
element is of this order.

From this consideration it is found that
the circulating atomic spin is the source current of neutrino pair emission
and the relevant current is given by
\begin{eqnarray}
&&
J_{eg}^{\alpha}(x) =  S^{\alpha} \frac{1}{\sqrt{\gamma}} \int dt  \rho_{eg}(t)\, \delta^{(4)} ( x - x_A(t)\,)
\,, 
\label {ion current}
\\ &&
H_w = \int d^3x \frac{G_F}{\sqrt{2}}  J_{eg}^{\beta}(x)\cdot \sum_{i,j=1,2,3} 
C_{ij}
\nu_i^{\dagger}(x) \sigma_{\beta} \nu_j(x)
\,,
\label {weak pair emission from excited atom}
\end{eqnarray}
where $x_A(t) = (t, \vec{r}_A(t)\,)$
 (written in terms of the time in the laboratory frame) is the  trajectory function
of excited ion in circular motion given by
\begin{eqnarray}
&&
\vec{r}_A(t) = \rho 
\left( \sin \frac{vt}{\rho}\,,  1-\cos \frac{vt}{\rho} \,, 0
\right)
\,,
\end{eqnarray}
where $\rho $ is the radius of the circular orbit.
The factor $1/\sqrt{\gamma}$ in eq.(\ref{ion current}) arises from the overlap of CM wave functions,
$\int d^3 X|  \Psi_i (X)|^2 = 1/\gamma$.

We adopt the interaction picture in which the kinetic and the mass terms
of neutrinos are taken as the free part of hamiltonian $H_0$, consisting
of diagonal terms of $b^{\dagger} b, d^{\dagger} d$ where
$b,d$ are annihilation operators of neutrino and anti-neutrino (in the Majorana
neutrino case $d= b$) when they are mode-decomposed using plane waves of
definite helicities.
The Fermi interaction (\ref {weak pair emission from excited atom}) due to the circular ion motion
gives rise to off-diagonal terms, in particular,
terms of the form, $bd, b^{\dagger} d^{\dagger}$.
In the perturbative picture this means that neutrino pairs may
be created at ion synchrotron.

The amplitude $ {\cal A}_{ij}(p_1h_1, p_2h_2; t)$ of neutrino-pair production of
momentum $\vec{p}_i$ (its energy given by
$E_i=\sqrt{p_i^2+m_i^2}$ for neutrino of mass $m_i$) and helicity $h_i$ time-evolves according to
\begin{eqnarray}
&&
i \partial_t {\cal A}_{ij}(p_1h_1, p_2h_2; t) =
i \partial_t \langle 0| d_i (p_2h_2; t) b_j (p_1h_1; t)| 0\rangle
=  \langle 0| [d_i( p_2h_2; t) b_j(p_1h_1; t), H_w]| 0\rangle
\,,
\end{eqnarray}
reducing the calculation to the commutator between the neutrino bilinear field $db$ and
the weak hamiltonian $H_w$.
The result for neutrino pair emission of a single flavor is given by a time integral,
\begin{eqnarray}
&&
{\cal A}_{ij}(p_1h_1, p_2h_2; t) = 
- i\sqrt{2} G_F \frac{1}{\sqrt{\gamma}}C_{ij}
\int_{-\infty}^t dt' e^{i(E+ E') t'} \tilde{J}_A^{\dagger}(\vec{p}_1 + \vec{p}_2; t') \cdot
j_{\nu}
\,,
\label {time integral form of amp}
\\ &&
\tilde{J}_A^{\alpha}(\vec{P}; t) =   \rho_{eg}(t) S^{\alpha}  e^{-i \vec{P}\cdot \vec{r_A}(t)}
\,, \hspace{0.5cm} 
j_{\nu} = u^{\dagger}(p_1h_1) \sigma v( p_2h_2) 
\,.
\end{eqnarray}
Here $u, v$ are associated plane-wave solutions of emitted neutrinos.

The basic interaction hamiltonian (\ref{weak pair emission from excited atom})
indicates a number of striking features of neutrino pair emission process.
Notably, it predicts a coherent (namely endowed with a definite phase relation
among two neutrinos in the pair) mixture of all neutrinos and anti-neutrinos
of three flavors.

The semi-classical approximation in the present work
is limited to the neutrino energy region in which ion recoil may 
be ignored, which allows the GeV region neutrino production
since the circulating ion energy is much larger.

\vspace{0.5cm} 
\section
{\bf Phase integral}

Pair emission rate defined by $P_{ij}(t; p_1h_1, p_2h_2) = \partial_t |{\cal A}_{ij}(p_1h_1, p_2h_2; t) |^2$
is given by
\begin{eqnarray}
&&
P_{ij}(t_0; p_1h_1, p_2h_2) =  4  G_F^2 |\rho_{eg}(0)|^2 \frac{1}{\gamma}
 |C_{ij} |^2
\int_{-\infty}^{0} dt S^{\alpha}\Re 
\left( {\cal N}_{\alpha \beta}( p_1h_1, p_2h_2)
e^{i(\Delta(0) - \Delta(t)\,)} \right) S^{\beta}
\nonumber 
\\ &&
=  4  G_F^2|\rho_{eg}(0)|^2 \frac{1}{\gamma} |C_{ij} |^2
\int_{0}^{\infty} dt S^{\alpha} \Re 
\left( {\cal N}_{\alpha \beta}( p_1h_1, p_2h_2)
e^{i(\Delta(0) - \Delta( - t)\,)} \right) S^{\beta}
\,,
\label {rate integral}
\\ &&
\Delta(t) = (E_1 + E_2 -  \frac{\epsilon_{eg}}{\gamma})t - (\vec{p}_1 + \vec{p}_2)\cdot \vec{r}_A(t)
\,, 
\\ && 
{\cal N}^{\alpha \beta}( p_1h_1, p_2h_2) = j_{\nu}^{\alpha}( p_1h_1, p_2h_2)
 (j_{\nu}^{\dagger})^{\beta}( p_1h_1, p_2h_2)
\,,
\end{eqnarray}
by taking an infinite time limit, which effective means that  time
contributing to the integral (\ref{rate integral}) 
is much larger than a small fraction of the orbital period $\rho/c$.
Explicit forms of ${\cal N}_{\alpha \beta}( p_1h_1, p_2h_2) $ 
may be evaluated by using formulas given in \cite{my-prd 07}.

\begin{figure*}[htbp]
 \begin{center}
 \epsfxsize=0.5\textwidth
 \centerline{\epsfbox{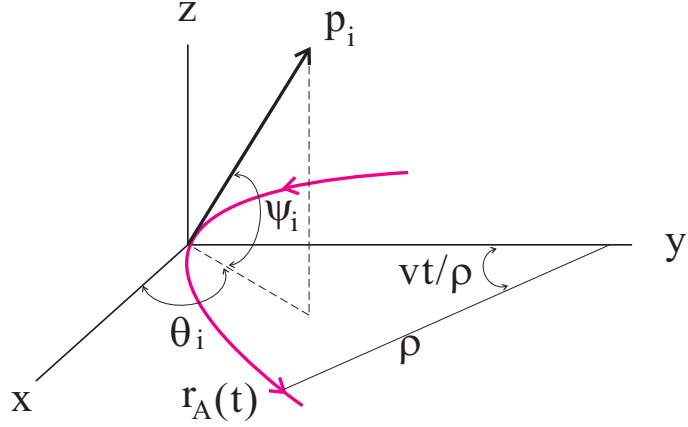}} \hspace*{\fill}\vspace*{2cm}
   \caption{Coordinate system for calculation of the phase integral.
Observation is made at a point far away on the (positive side of) x-axis.
Circular motion of excited ion in the $(x,y)$ plane is depicted in red.
The angle $\psi_i$ is defined to be zero in the ion orbit plane,
while $\theta_i$ is the angle measured from the tangential direction to
the ion beam, with the negative region $\theta_i < 0$ being defined 
towards the inner region of the circular orbit.
}
   \label {coordinate system}
 \end{center} 
\end{figure*}

The important phase factor in the integral  is given by
\begin{eqnarray}
&&
\Delta (0) - \Delta(-t)
= (E_1 + E_2 - \frac{\epsilon_{eg}}{\gamma}+ i \frac{1}{\gamma T_2} )t -\rho \left(
(p_1+ p_2)_x \sin \frac{vt}{\rho} +  (p_1+ p_2)_y (1 - \cos\frac{vt}{\rho} \,)
\right)
\,.
\label {phase factor}
\end{eqnarray}
Let us introduce directional angles of emitted neutrinos:
\begin{eqnarray}
&&
\vec{p}_i = p_i (\cos \psi_i \cos (\theta_i + \frac{vt}{\rho}) , \cos \psi_i \sin  (\theta_i + \frac{vt}{\rho}), \sin \psi_i)
\,, \hspace{0.5cm}
- \frac{\pi}{2} \leq \psi_i \leq \frac{\pi}{2} 
\,, \hspace{0.5cm}
- \pi \leq \theta_i \leq \pi 
\,.
\end{eqnarray}
Angles are measured at an observation point away from the circular motion.
The forward and the background directions with respect to the ion beam
correspond to $| \theta_i | < \pi/2$ and $| \theta_i | > \pi/2$, respectively.
See Fig(\ref{coordinate system}) for this coordinate system.

The phase factor $\Delta (0) - \Delta(-t)$ of eq.(\ref{phase factor}) 
contains three terms: in addition to the main term $\propto E_1+E_2$,
one is from the circulating ion proportional to $\rho$, the radius of the orbit,
and the other is proportional to the level spacing $\epsilon_{eg}$.
Under  the normal condition  one may ignore
the imaginary component $\propto 1/T_2$, since $\epsilon_{eg} \gg 1/T_2$.
The most important observation in the present work
is that an input of de-excitation energy $\epsilon_{eg}$
may lead to cancellation of three terms and to existence of stationary phase points 
in the relevant phase integral along the real axis of time.
On the other hand,
without the $\epsilon_{eg}$ term one can show that the phase is positive definite.
As is well known in mathematical physics, contribution around  stationary points
does not suffer from large suppression unlike constructive interference contributing with
the same sign phase.
This was the case without the $\epsilon_{eg}$ term such as synchrotron radiation and 
neutrino pair emission from the ground state ion.
The well-known exponential cutoff arising from the constructive interference
gives rise to the cutoff energy
of emitted photon $\approx \gamma^3/\rho$ in synchrotron radiation \cite{schwinger}.
This cutoff also occurs for neutrino pair emission at electron synchrotron,
restricting available neutrino energies up to a keV range.

The crucial condition for the presence of stationary points is derived by
setting vanishing time derivative of eq.(\ref {phase factor}), leading to an equality,
\begin{eqnarray}
&&
E_1 + E_2 - \frac{\epsilon_{eg}}{\gamma} -  v \sum_i p_i\cos \psi_i \cos(\theta_i +\frac{vt}{\rho}) = 0
\,.
\label {stationary point condition}
\end{eqnarray}
Infinitely many stationary points exist along
the real axis of time $t$.
It turns out that the most important contribution comes from the 
point nearest to $t=0$, the end point of the integration range $t \geq 0$.
Note that without the $\epsilon_{eg} $ term there is no stationary point solution on the real axis,
hence the possibility of large neutrino pair production
at electron synchrotron is excluded.

Let us consider the in-plane forward direction of $\psi_i = \theta_i = 0$ for  $E_i \gg \epsilon_{eg}$.
One may expand the left hand side function
of eq.(\ref{stationary point condition}) in powers of time variable $t$, to solve the stationary point condition.
The look-back time of the stationary point $t=t_c$ is then derived as
\begin{eqnarray}
&&
t_c \sim  \rho \sqrt{\frac{2\epsilon_{eg}}{\gamma(E_1 + E_2) } - \frac{1}{\gamma^2}}
\,,
\end{eqnarray}
for $E_1 + E_2 < 2 \epsilon_{eg} \gamma$.
In the following we shall take the kinematical region in which this stationary point exists.
In order to evaluate the phase integral to a good approximation, we use the power series expansion, 
to derive
\begin{eqnarray}
&&
x = 
 \frac{|A|}{\sqrt{2}\rho}
\left( t
+ (1 - \frac{1}{\gamma^2})^{-1/2}\frac{ p_1 \cos \psi_1 \sin \theta_1 + p_2 \cos \psi_2 \sin \theta_2}
{ p_1 \cos \psi_1 \cos \theta_1 + p_2 \cos \psi_2 \cos \theta_2}\rho
\right) 
\,, 
\\ &&
\Delta(0) - \Delta( - t)  \sim \xi \left( - \frac{3}{2}x + \frac{1}{2} x^3 \right)
\,,
\\ &&
\xi = \frac{2\sqrt{2}}{3} \frac{\rho}{|A|} 
\left(
\frac{\epsilon_{eg}}{\gamma} 
 -E_1 - E_2   +  (1 - \frac{1}{\gamma^2})^{1/2} (p_1 \cos \psi_1 \cos \theta_1 + p_2 \cos \psi_2 \cos \theta_2)
+ B
\right)
\,,
\\ &&
|A|^2 =   (1- \frac{1}{\gamma^2})^{1/2} 
\left( \frac{p_1 \cos \psi_1 \cos \theta_1 + p_2 \cos \psi_2 \cos \theta_2 }
{ \frac{\epsilon_{eg}}{\gamma}  
-E_1 - E_2  +  (1 - \frac{1}{\gamma^2})^{1/2} (p_1 \cos \psi_1 \cos \theta_1 + p_2 \cos \psi_2 \cos \theta_2)
+ B}
\right)
\,, 
\\ &&
B = \frac{1}{2}  (1 - \frac{1}{\gamma^2})^{1/2} \frac{(p_1 \cos \psi_1 \sin \theta_1 + p_2 \cos \psi_2 \sin \theta_2)^2 }
{p_1 \cos \psi_1 \cos \theta_1 + p_2 \cos \psi_2 \cos \theta_2 }
\,.
\end{eqnarray}
In deriving this equation, we shifted the integration variable $t$ such that
$O(t^2)$ terms are  eliminated.
The power series expansion in terms of time $t$
has been retained up to 3rd order of $t^3$,
because still higher order terms 
are suppressed by powers of $t_c/\rho$.

The condition that the stationary point is within the integration range $t \geq 0$ gives
a limitation of emitted angles.
An angular region deep inside the circle of ionic motion 
gives stationary points in the forbidden region of $t < 0$,
hence does not give large neutrino pair emission rates.
This forbidden region is defined by
\begin{eqnarray}
&&
\sqrt{2} (E_1 \theta_1 + E_2 \theta_2 ) < - 
\sqrt{( \frac{\epsilon_{eg}}{\gamma} 
 - \frac{1}{2\gamma^2} (E_1 + E_2)-\frac{1}{2}(E_1 (\theta_1^2 + \psi_1^2) + 
E_2 (\theta_2^2 + \psi_2^2)\, )(E_1 + E_2)}
\,.
\end{eqnarray}

The necessary phase integral of $x-$variable involves 
a smoothly varying function $h$ of time and it has a form,
\begin{eqnarray}
&&
\int_0^{\infty} dx h(x) \cos \xi (\frac{1}{2} x^3 - \frac{3}{2}x ) \sim
h(1) \frac{\pi}{3} \left( J_{1/3} (\xi) + J_{-1/3} (\xi) \right)
\,,
\label {phase integral 0}
\end{eqnarray}
where $J_{\nu}(z)$ is the Bessel function, and
$h(x)$ is a smoothly varying function of time given by squared matrix element
of neutrino pair emission.
The large radius limit of $\rho \rightarrow \infty$, hence the 
$\xi (\propto \rho) \rightarrow \infty$,  is important for calculation
of differential rates, since the radius $\rho$ is much larger than any
microscopic length scale involved.
The limit gives
\begin{eqnarray}
&&
\int_0^{\infty} dx h(x) \cos \xi (\frac{1}{2} x^3 - \frac{3}{2}x ) 
\rightarrow 
\sqrt{\frac{2\pi}{3}} \cos (\xi - \frac{\pi}{4}) \frac{ h(1)}{\sqrt{\xi}} 
\,,
\label {phase integral h}
\end{eqnarray}
as $\xi \rightarrow \infty$.
This asymptotic formula may also be derived directly using
the steepest descent, or the stationary phase method of mathematical physics.
The stationary point appears at $x= 1$, which implies that $t_c =  \sqrt{2} \rho/|A|$.
In addition to the phase factor given here there is a constant phase factor 
arising from the phase at the stationary point, namely $\Delta(0) - \Delta(-t_c)$, 
which however gives a negligible contribution.

The fact that the phase integral proportional to rate, (\ref{phase integral 0}) or  (\ref{phase integral h}),
can give negative values for some value of $\xi$ might appear odd.
But since this is time derivative of a positive quantity (probability), this may occur
without any violation of fundamental principles.
Indeed, this also occurs in the usual formula of synchrotron radiation
\cite{schwinger}.
The quantity $\xi$ is actually a complicated function of neutrino energies, their emission
angles, the boost factor $\gamma$, and the atomic energy scale $\epsilon_{eg}$.
The region of these variables that effectively contributes with a large rate
is found to give mostly positive rates.
Thus, there is no serious problem of  the negative rate.
An alternative method used in the case of synchrotron radiation \cite{schwinger}
treats the time and one of the angular variables,
$\theta$ (essentially not measurable) symmetrically in integration
by changing integration variables in a clever way.
A generalization of this method to the case of neutrino pair emission
might be possible with much effort.

For a finite value of $T_2$, the stationary point moves to a point slightly off the
real axis, introducing a small correction
to $\epsilon_{eg} $ replaced by $ \epsilon_{eg} - i/T_2$.
The effect of this shift is small.

It would be instructive, before proceeding,  to mention the limiting case of
$\epsilon_{eg} \rightarrow 0$ in our problem.
In the limit  the stationary point $t_c$ approaches 
the end point of time integration range and the phase space of neutrino momenta
shrinks to zero.
Thus, rate due to the mechanism considered vanishes in the limit.

As another extension 
we would like to mention other contributions than the spin current contribution
considered here.
Contributions from the excited state and the ground state are
proportional to $\rho_{ee}, \rho_{gg}$ which do not have the $\epsilon_{eg}$ factor
in the phase,
hence this case too has no stationary point on the real time axis.
 Result of the phase integral in these cases is given in terms of
the modified Bessel function much like in the synchrotron radiation.
The neutrino energy spectrum then suffers from the exponential cutoff
of order $ \gamma^3/ \rho$, which is typically in the keV region.
Since the weak interaction rates scales with $ {\rm energy}^5$,
rates are negligibly small.
We have neglected these contributions.

\vspace{0.5cm} 
\section
{\bf Differential emission rate of a single  pair}

Let us first write down the squared spin amplitude (\ref {helcity summed spin factor})
in our coordinate system:
\begin{eqnarray}
&&
{\cal M} = \frac{1}{3} \gamma^2 S_e^2
\left(
1+ \frac{1}{3} \cos \psi_1 \cos \psi_2 \cos (\theta_1 - \theta_2) + \frac{1}{3} \sin\psi_1 \sin \psi_2
\right)
\,.
\end{eqnarray}
For simplicity we took neutrinos to be massless which is adequate for our purpose.
Effects of finite neutrino masses are significant only for 
$E_i < O(m \gamma)$, which is of order keV for
$m = 0.1 {\rm eV}\,, \gamma= 10^4$.
Since neutrino pair production rates are small for this energy range,
we shall ignore effects of finite neutrino masses for discussion of
production rates.

A straightforward calculation using this result gives the differential production rate for a
neutrino pair  $\nu_i \bar{\nu}_j$ of mass eigenstates.
In order to avoid complication, we shall write down this formula
in the leading approximation of large boost factor;
\begin{eqnarray}
&&
\frac{d^4 \Gamma_{ij} }{dE_1 dE_2 d\Omega_1d \Omega_2}
= \frac{4 G_F^2}{ 2^{7/4}\cdot 3\sqrt{3\pi}(2\pi)^6}  |C_{ij} |^2
S_e^2N|\rho_{eg}(0)|^2 \gamma \sqrt{\rho}
 E_1^2   E_2^2 F^{-1/4}
\,,
\label {differential spectral rate h}
\\ &&
F = (E_1 + E_2) (\frac{\epsilon_{eg}} {\gamma} - \frac{E_1+E_2}{2\gamma^2})
- \frac{1}{2}(E_1^2 \psi_1^2 + E_2^2 \psi_2^2) -\frac{E_1E_2}{2}(\theta_1- \theta_2)^2
- \frac{\epsilon_{eg}}{2\gamma} ( E_1\theta_1^2 +  E_2\theta_2^2)
\,.
\label {differential spectral rate h2}
\end{eqnarray}
The spin factor is given by ${\cal M} \sim 4 \gamma^2 S_e^2/9 $ in this approximation.
The function $F$ is more complicated in the most general case of the boost factor,
which may be inferred from eq.(\ref{double rate for E1g}) for (the electric dipole) photon emission.
There is a constraint on angle factors given by $F \geq 0$.
This constraint gives
angular restriction worked out for apertures (given for simplicity to the case $E_1=E_2 = E$),
\begin{eqnarray}
&&
\Delta \psi = O\left(\frac{1}{\gamma} \sqrt{\frac{2(E_m - 2E) }{ E}} \right)
\,, \hspace{0.5cm}
\Delta \theta = O\left(\sqrt{\frac{ E_m - 2E} { E_m }} \right) 
\,, \hspace{0.5cm}
E_m = 2\gamma \epsilon_{eg}
\,.
\label {angular range}
\end{eqnarray}
While $\Delta \theta_i$ is of order unity individually,
the opening angle of two neutrinos of the pair is limited by
\begin{eqnarray}
&&
\Delta |\theta_1 - \theta_2 | < O\left(\frac{1}{\gamma} \sqrt{ \frac{(E_1+E_2) (E_m - E_1-E_2) }{E_1 E_2 }} \right)
\,.
\end{eqnarray}
The suppression by $1/\gamma$ for the opening angle is of great interest from
the point of oscillation experiments, since it leaves open for the possibility of
a coherent neutrino pair interaction at measurement sites.
We shall have much to discuss in the next section.

\begin{figure*}[htbp]
 \begin{center}
 \epsfxsize=0.5\textwidth
 \centerline{\epsfbox{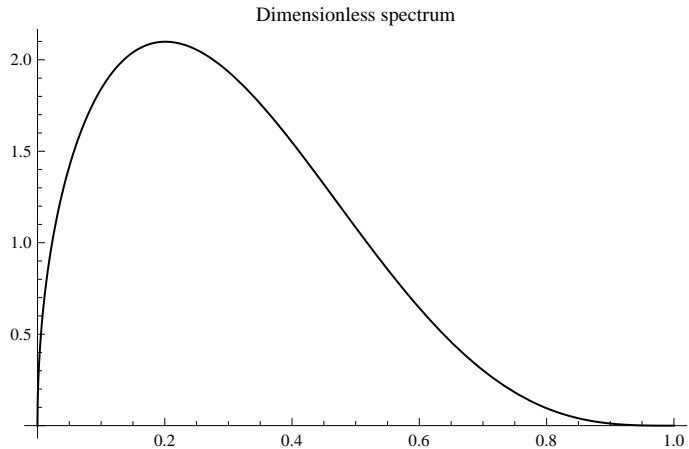}} \hspace*{\fill}
   \caption{Normalized universal spectrum function $f(x)$ of eq.(\ref{universal spectrum})
divided by its total integral $\int_0^1 dy f(y)$ in which $x= E/E_m, E_m = 2 \epsilon_{eg} \gamma$
is the fractional neutrino energy.
}
   \label {universal function}
 \end{center} 
\end{figure*}

Integration over four angle factors, with $d\Omega_i = \cos \psi_i d\psi_i d\theta_i$, may be carried out, 
to give
\begin{eqnarray}
&&
\int d\Omega_1 \int d\Omega_2 F^{-1/4}
\sim \frac{V_4}{14} (\frac{\epsilon_{eg}}{\gamma})^{5/4}
\frac{ (E_1 + E_2)^{5/4} } {(E_1 E_2)^{3/2} }
(1 - \frac{E_1 + E_2}{E_m})^{7/4}
\,,
\end{eqnarray}
where $V_4 = \pi^2/2$ is the volume of 4-dimensional sphere of unit radius.
Using this result, the double differential energy spectrum becomes
\begin{eqnarray}
&&
\hspace*{-1cm}
\frac{d^2 \Gamma_{ij}}{dE_1 dE_2} = \frac{1}{21 \cdot 2^7\cdot 2^{3/4} \sqrt{3\pi} \pi^4 }
 |C_{ij} |^2 S_e^2N|\rho_{eg}(0)|^2  \sqrt{\rho} \gamma
 (\frac{\epsilon_{eg}}{\gamma})^{5/4} G_F^2 (E_1 E_2)^{1/2} (E_1 + E_2)^{5/4}
(1 - \frac{E_1 + E_2}{E_m})^{7/4}
\,. 
\nonumber
\\ &&
\end{eqnarray}
The relation $\sum_j |C_{ij}|^2 = 1/4$ was used.
Further integration gives the single neutrino energy spectrum and the total
pair production rate:
\begin{eqnarray}
&&
\frac{d\Gamma_i}{dE} = \frac{1}{21 \cdot 2^{10}\cdot \sqrt{6\pi}\pi^4 }
 S_e^2N|\rho_{eg}(0)|^2  \sqrt{\rho \epsilon_{eg} } G_F^2 E_m^4 \frac{1}{\gamma}
f(\frac{E}{E_m})
\,,
\\ &&
f(x) = \sqrt{x} \int_0^{1-x} dy y^{1/2} ( y+ x)^{5/4} ( 1- x - y)^{7/4}
\,, \hspace{0.5cm}
\int_0^1 dx f(x)  \sim 0.00 727
\,,
\label {universal spectrum}
\\ &&
\Gamma_i \sim 0.00 73 \frac{1}{21 \cdot 2^{10}\cdot \sqrt{6\pi}\pi^4 \gamma}
 S_e^2 N |\rho_{eg}(0)|^2 \sqrt{\rho  \epsilon_{eg}} G_F^2 E_m^5
\,.
\label {total rate}
\end{eqnarray}
Production rate does not depend on the mass eigenstate label,
since we assumed the massless neutrino for this calculation.
The normalized universal spectrum function $f(x)/\int_0^1 dy f(y)$
is plotted in Fig(\ref {universal function}).
The average neutrino energy is $\sim 0.30 E_m$.
The end point behavior at $x=1$ gives the threshold behavior
$\propto (E_m - E)^{13/4}\,, E_m = 2 \epsilon_{eg} \gamma$ at the highest neutrino energy
and $\propto \sqrt{E}$ in the infrared region of $E \rightarrow 0$.

Dependence of the total rate on  parameters
$\epsilon_{eg}\,, \gamma$ taken as independent is $\propto \gamma^{4} \epsilon_{eg}^{11/2}  $.
Along with the relation $E_m =2 \epsilon_{eg}\gamma$,
we conclude that
it is desirable to choose highly stripped heavy ions in order to
achieve  both high energy neutrino and large production rates.
Rates further depend on $N |\rho_{eg}(0)|^2$ of injected ion,
which requires a coherence of large $\rho_{eg}(0)$.
We shall derive a constraint on this coherence factor
in Section 6.
A numerical estimate then gives
\begin{eqnarray}
&&
\Gamma = \sum_i \Gamma_i \sim 3.1 \times 10^{21} {\rm Hz} (\frac{\rho}{ 4 {\rm km}})^{1/2} 
\frac{ S_e^2 N |\rho_{eg}(0)|^2}{10^8}
(\frac{\gamma}{10^4})^4 (\frac{\epsilon_{eg}}{50 {\rm keV}})^{11/2}
\,,
\label {total rate}
\\ &&
{\rm with }\;
E_m = 2\epsilon_{eg} \gamma = 1{\rm GeV}  \frac{\epsilon_{eg}}{50 {\rm keV}} \frac{\gamma}{10^4}
\,.
\end{eqnarray}

We may offer an interpretation of
dependence of the total rate
on involved various quantities.
Ignoring dimensionless numerical values one  has the relation,
\begin{eqnarray}
&&
\Gamma \propto \frac{1}{\gamma} \cdot
N|\rho_{eg}(0)|^2 \cdot G_F^2 E_m^5 \cdot
\sqrt{ \rho \epsilon_{eg}}
\,.
\end{eqnarray}
Each factor written here has a clear meaning.
What this dependence implies is a scaling law
with the boosted factor $\gamma$ 
in the laboratory frame of the circular motion.
Note first that there is a hidden $\gamma$ factor
in the radius of $1/\rho =Q eB/(\gamma M_A)$
with $M_A$ the ion mass and $Qe$ the charge of ion.
Except for the first factor $1/\gamma$ which arises
from the prolonged lifetime $\propto \gamma$,
other factors are dictated by the simple scaling of the basic atomic
energy, with $\epsilon_{eg} \rightarrow \gamma \epsilon_{eg}$.
This scaling law holds in the photon emission rate
discussed in Section 6.

\begin{figure*}[htbp]
 \begin{center}
 \epsfxsize=0.6\textwidth
 \centerline{\epsfbox{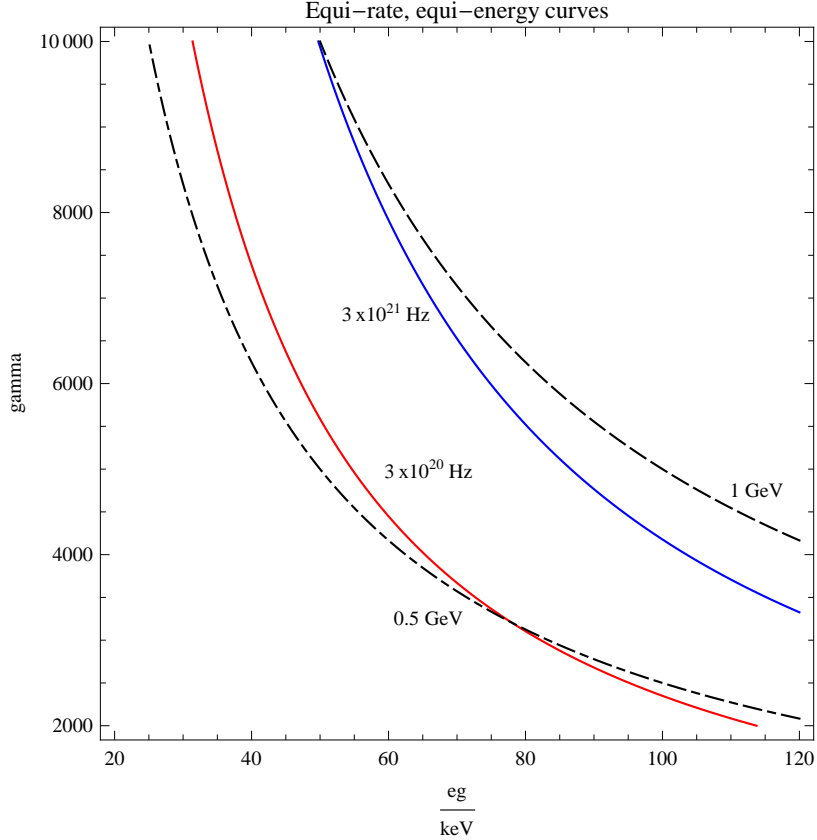}} \hspace*{\fill}
   \caption{ 
Total equi-rate (given by eq.(\ref{total rate})) curves for
$3\times 10^{20},  3\times10^{21}$ Hz's in colored solid 
in ($\epsilon_{eg}/{\rm keV}, \gamma$) plane.
Also plotted is $E_m = 1$GeV curve in dashed black and  $E_m = 0.5$GeV curve in dashed-dotted black. 
$N |\rho_{eg}(0)|^2 S_e^2= 10^{8}$ is assumed.
The normal hierarchical mass pattern of the smallest neutrino mass 0 is taken, with
$\epsilon_{eg} = 50 {\rm keV}, \gamma = 10^4$ the synchrotron radius, $ \rho=4 {\rm km}$.
}
   \label {total rate contour}
 \end{center} 
\end{figure*}

For a variety of expected neutrino experiments based on
CP-even neutrino beam,
it is important to have a beam of neutrino energy high enough in
the GeV region (at minimum,  larger than O(200) MeV), since only then one can clearly detect
the charged current (CC) interaction of $\nu_{\mu}, \bar{\nu}_{\mu}$.
If this requirement is not fulfilled, one only has
CC interaction of $\nu_e, \bar{\nu}_e$ and all kinds of
neutral current (NC) interaction including $\nu_{\tau}, \bar{\nu}_{\tau}$.
NC process has a lower rate and experiments are harder.
The GeV neutrino production requires $2 \epsilon_{eg} \gamma \geq 1$GeV
for this combination of the ion parameter and the boost factor.
For $\gamma \leq 10^4$, it is necessary to have $\epsilon_{eg} \geq 50$keV
in order to reach 1 GeV neutrino energy.
It is then important to  excite electrons  deeply bound in 
highly stripped ions in order to reach the keV binding energy of valence electrons
in ion.
This might be a non-trivial problem, but we assume
that this is possible \cite{excitation of atoms}.
Judging from Fig(\ref{total rate contour}) it appears
that there is an excellent chance of neutrino  experiments
in  O(0.5 $\sim$ 1 GeV) energy range, which roughly gives a large rate
of order, $10^{20} \sim 10^{21}  $Hz
of the neutrino pair emission.
A large value of $\epsilon_{eg}$ and a large boost factor $\gamma$ are required
to accomplish this goal.

Very importantly as a caveat,
a new scheme of continuous injection or generation of coherent excited ion beam
should be invented, because an excited ion, once it produces 
the neutrino pair, is not expected to be reusable for
another source.
The design and realization of this scheme might be challenging, and one may have to tolerate
a sizable and unavoidable reduction of effectively usable ion number.

A possible problem of highly stripped heavy ions is
their large magnetic dipole (M1) transition rate
\cite{uetake}.
For concreteness let us take an example of
He-like ion , Pb$^{80+}$.
A good candidate for the initial ionic level is
$| e \rangle = \left( (2s)(1s)\right)^{3}_{J = 1} $
(a spin triplet state described in $jj$ coupling scheme)
of level spacing $\epsilon_{eg} \sim 70$keV.
The beam loss rate due to M1 photon emission
is
\begin{eqnarray}
&&
\Gamma_{\gamma} = \gamma_{M1} N \rho_{ee}(0)
\,,
\end{eqnarray}
where the M1 decay rate $ \gamma_{M1}$ is $\sim 3.4 \times 10^{13}$Hz
according to \cite{he-like m1}.
By requiring that this loss rate is smaller than
the neutrino pair emission rate, one derives
\begin{eqnarray}
&&
|\rho_{eg}(0)|^2 > O(0.1) \rho_{ee}(0) (\frac{\gamma }{10^4 })^{-4}
\,.
\end{eqnarray}
Assuming the relation
\(\:
|\rho_{eg}(0)|^2 =  \rho_{ee}(0)  \rho_{gg}(0)
\:\)
that holds for a quantum mixture of pure states,
one may further derive a lower bound for $\rho_{gg}(0)$
of order $0.1 (\gamma/10^4 )^{-4}$.
This condition in the general case of He-like ions, with the inequality 
$\rho_{gg}(0) < 1$,
further  gives constraint on a relation of 
the boost factor and the level spacing.
Details on these constraints should be worked out
after more detailed R and D investigation on candidate heavy ions.

\vspace{0.5cm}
\section
{\bf  High energy gamma ray beam}

For completeness we shall present main results
for high energy gamma emission from the circulating excited ion.
The high energy gamma ray emission occurs between different parity states
among which E1 transition is allowed.
The formalism in the main text is readily adapted to this case
and we shall be brief in presenting results.

The basic hamiltonian operator of E1 photon emission is 
 (using  a similar notation as in the text)
\begin{eqnarray}
&&
H_{\gamma} = \int d^3 x \frac{e}{m_e} \vec{A}(x)\cdot \vec{J}_{\gamma}(x)
\,, \hspace{0.5cm}
\vec{J}_{\gamma}(x) =  \frac{1}{\sqrt{\gamma}} \gamma \int dt \rho_{eg}(t)  \vec{p}_{eg}\delta ( x - x_A(t)\,)
\,,
\label {e1 emission hamiltonian}
\end{eqnarray}
where $\vec{A}(x) = e^{i\vec{k}\cdot \vec{x}} \vec{e}_{\vec{k}}/\sqrt{2\omega V}$ is the vector potential of emitted
plane-wave photon
($\vec{e}_{\vec{k}}$ being the polarization of photon).
The other contribution arising from the center of mass (CM) motion part of ion as a whole $\propto \vec{P}_A$
 (atomic momentum) has been omitted, because it does not contribute to the internal atomic transition of
$|e \rangle \rightarrow |g \rangle$.
The CM part  gives contribution similar to the usual electron's synchrotron radiation
and gives rates much smaller than the rest of contribution.
See more on this point.

The matrix element of the internal part $\propto \vec{p}_{eg}$ leading to eq.(\ref {e1 emission hamiltonian})
has been worked out as follows.
The relativistic form of interaction hamiltonian density after the Lorentz boost of
$\gamma$ factor is given by
\begin{eqnarray}
&&
e\gamma \int d^3 x \langle g |\left( 
\vec{A} \cdot \psi^{\dagger} \vec{\alpha} \psi + \vec{A} \cdot\vec{\beta} \psi^{\dagger}  \psi 
\right) | e \rangle
\,,
\end{eqnarray}
using the radiation gauge in the atomic rest frame,
where $\vec{\alpha}$ is the Dirac 4$\times$4 matrix, $\vec{\alpha} = \gamma_0 \vec{\gamma}$,
and $\vec{\beta} $ is the Lorentz boost vector.
The orthogonality of (non-relativistic) wave functions of $|e\rangle$ and $|g\rangle$ gives
the vanishing second contribution $\propto \psi^{\dagger}  \psi $ to the leading first order
to $v/c$ ($v$ being the velocity of atomic electron in its rest frame).
The first contribution gives the internal contribution of eq.(\ref{e1 emission hamiltonian})
when the matrix element $\int d^3x \langle g| e^{i\vec{k}\cdot \vec{x}} \psi^{\dagger} \vec{\alpha} \psi |e \rangle $ 
is written in the atomic rest frame, taking the long wavelength approximation of $\vec{k} \rightarrow 0$
valid comparing with a larger inverse atomic length scale.
The atomic matrix element may further be recast into the usual dipole form, using the equation
of motion:
$\vec{p}_{eg}/m_e = - i\epsilon_{eg} \vec{r}_{eg}$ with $e \vec{r}_{eg}$ 
the dipole matrix element for E1 transition.

Calculation of the phase integral involves the energy-momentum 
$(\omega, \vec{k})\,, |\vec{k}| = \omega$ of a single  photon.
Stationary phase points appear due to the presence of the energy $\epsilon_{eg}$
in the phase integral.
Straightforward calculations using the same approximation as in the text lead to photon emission rates.
It would be instructive to start from a detailed discussion of the angular distribution.
The double differential emission rate is given, to the best accuracy we know of,
by
\begin{eqnarray}
&&
\hspace*{-1cm}
\frac{d^2 \Gamma}{d\omega d\Omega} =
\frac{1}{2^{1/4} 16 \pi^3} N |\rho_{eg}(0)|^2 \gamma \sqrt{\rho} \frac{\gamma_{eg}}{\epsilon_{eg}}
\omega^{3/4} 
\left( 
\cos \theta \cos \psi (\frac{\epsilon_{eg}}{\gamma} - \omega +
\frac{1}{2} (1- \frac{1}{\gamma^2})^{1/2}\omega
\frac{1+\cos^2\theta}{\cos \theta} \cos \psi )
\right)^{-1/4}
\,,
\label {double rate for E1g}
\end{eqnarray}
where the squared dipole moment $\vec{r}_{eg}^{\,2}$ was replaced by the
related decay rate $\gamma_{eg}$ (Einstein's A-coefficient).
The bracketed quantity in the argument of the negative fractional power $-1/4$ must be positive definite.
To the leading order of the boost factor $\gamma$, this quantity is approximated near the
forward direction by
\begin{eqnarray}
&&
\frac{\epsilon_{eg}}{\gamma} - \frac{\omega}{2\gamma^2} 
- \frac{\omega}{2} \psi^2
- \frac{\epsilon_{eg} }{2\gamma} \theta^2
\,.
\end{eqnarray}
The positivity requires the inside region of an ellipsoid in $(\theta, \psi)$ plane,
\begin{eqnarray}
&&
\psi^2 + 
\frac{\epsilon_{eg} }{ \gamma\omega} \theta^2  \leq \frac{\omega_m  - \omega} {\gamma^2 \omega }
\,, \hspace{0.5cm}
\omega_m = 2\gamma\epsilon_{eg}
\,,
\end{eqnarray}
along with $\omega \leq  \omega_m$.
Thus, there exists an interesting angular asymmetry:
the cylindrical symmetry around the tangential direction is broken.
With this approximation, the double differential rate becomes
\begin{eqnarray}
&&
\frac{d^2 \Gamma}{d\omega d\Omega}  \sim
\frac{1}{2^{1/4}\cdot 16 \pi^3} N |\rho_{eg}(0)|^2 \gamma \sqrt{\rho} \frac{\gamma_{eg}}{\epsilon_{eg}}
\omega^{3/4} 
\left( 
\frac{\epsilon_{eg}}{\gamma} - \frac{\omega}{2\gamma^2} 
- \frac{\omega}{2} \psi^2
- \frac{\epsilon_{eg} }{2\gamma} \theta^2
\right)^{-1/4}
\,.
\end{eqnarray}
The small angle approximation here is valid only for
a small value of $2( \omega_m - \omega)/(\gamma^2 \omega)$.
The approximation clearly breaks down at the infrared limit $\omega \rightarrow 0$.

\begin{figure*}[htbp]
 \begin{center}
 \epsfxsize=0.5\textwidth
 \centerline{\epsfbox{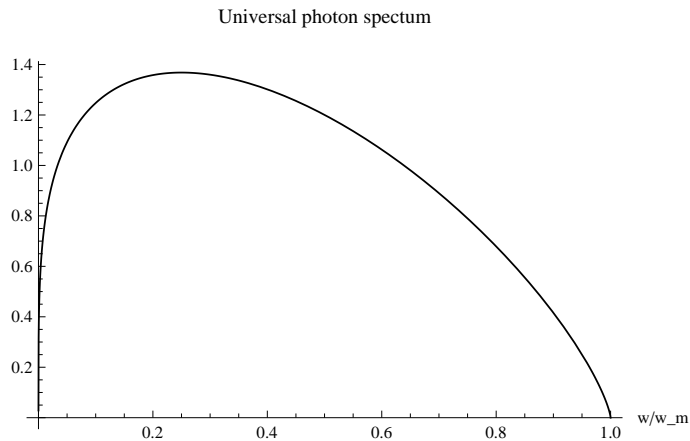}} \hspace*{\fill}
   \caption{Normalized universal energy spectrum of 
photons emitted from excited ions.
}
   \label {high gamma 2}
 \end{center} 
\end{figure*}

Further angular integration is straightforward if one uses the small angle approximation, leading to
the photon energy spectrum and finally the total emission rate,
\begin{eqnarray}
&&
\frac{d\Gamma}{d\omega} = 
\frac{1}{24\pi^2} N |\rho_{eg}(0)|^2 \gamma  \frac{\gamma_{eg}}{\epsilon_{eg}}
\sqrt{\rho \epsilon_{eg}} (\frac{\omega}{ \omega_m} )^{1/4}
 (1 -\frac{\omega}{\omega_m} )^{3/4} 
\,.
\\ &&
\Gamma = \frac{ I}{12\pi^2} N |\rho_{eg}(0)|^2 \gamma^{2} 
\gamma_{eg}
\sqrt{\rho \epsilon_{eg}} 
\,, \hspace{0.5cm}
I  = \int_0^1 dy y^{1/4} (1-y)^{3/4} \sim 0.4165
\,.
\end{eqnarray}
The dimensionless spectrum function $x^{1/4} (1-x)^{3/4}/I\,, x = \omega/\omega_m$
 is plotted in Fig(\ref{high gamma 2}) after renormalization.
Its end point is at $\omega_m = 2\gamma \epsilon_{eg}$ and
the averaged energy value is $0.42 \omega_m$.
A typical value of the total photon emission rate is
\begin{eqnarray}
&&
\Gamma \sim 1.1 \times 10^{29} {\rm Hz}\,
\frac{\gamma_{eg} }{100 {\rm MHz}} \sqrt{\frac{\rho}{4 {\rm km}}} 
(\frac{\epsilon_{eg}}{50 {\rm keV}})^{1/2}  (\frac{\gamma}{10^4})^{2}
\frac{ N |\rho_{eg}(0)|^2}{10^8}
\,,
\end{eqnarray}
for $\omega_m = $ 1GeV case.

In a similar fashion
as for the neutrino pair emission,
the formula of the total E1 photon emission rate may
be interpreted using the $\gamma-$scaling law.
Let us ignore dimensionless numerical factors for this purpose.
The total rate then has dependence on various quantities:
\begin{eqnarray}
&&
\Gamma \propto 
\frac{1}{\gamma} \cdot N|\rho_{eg}(0)|^2 \cdot 
e^2 \vec{r}_{eg}\,^2 \omega_m^3 \cdot
\sqrt{\rho \epsilon_{eg}}
\,, \hspace{0.5cm}
\omega_m = 2\gamma \epsilon_{eg}
\,.
\end{eqnarray}
When one regards the atomic dipole $e \vec{r}_{eg}$ as an invariant
and intrinsic quantity to atom,
other energy factor scales as $\propto \gamma $
under the Lorentz transformation,
along with the prolonged lifetime factor $1/\gamma$ in front.
This law explains $\gamma$ and $\epsilon_{eg}$ dependence
of photon emission rate $\propto \gamma^2 \epsilon_{eg}^{7/2}$,
as well as that of neutrino pair emission $\propto \gamma^4 \epsilon_{eg}^{11/2}$
($\rho$ regarded as $\gamma$ independent).

Comparison with the usual synchrotron radiation may be instructive and also
interesting.
We can work out photon emission caused by ion circular motion in the ground state
(or kept in the excited state)
or simply an electron's circular motion, using the same calculation technique as above.
The basic hamiltonian arises from the omitted $\propto \vec{P}_A$
term and the calculation is purely classical unlike the semi-classical approximation in the case of
photon emission from an excited level.
There is no lifetime related factor $\propto 1/\gamma$ in this case,
because the synchrotron emission is not a decay process.
The result differs in an essential way from
the case of excited ion,  in that there is no stationary point of time integral.
The phase integral in this case takes the form,
\begin{eqnarray}
&&
\int_0^{\infty} dx h(x) \cos \xi (\frac{1}{2} x^3 + \frac{3}{2}x ) 
\rightarrow 
\sqrt{\frac{\pi}{6}} e^{-\xi} \frac{ h(0)}{\sqrt{\xi}} 
\,.
\label {phase integral h2}
\end{eqnarray}
There is no phase cancellation unlike in the case of
excited ion.
Instead,
the exponential cutoff emerges for large $\xi$.

\begin{figure*}[htbp]
 \begin{center}
 \epsfxsize=0.6\textwidth
 \centerline{\epsfbox{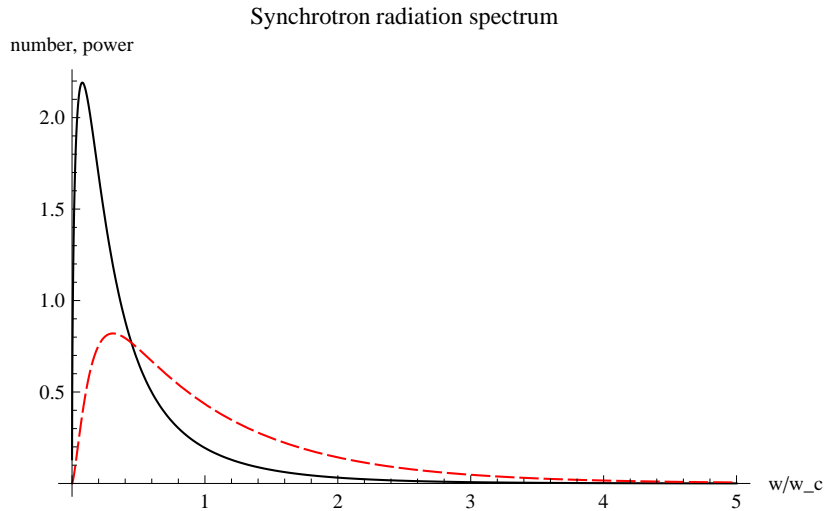}} \hspace*{\fill}
   \caption{Normalized synchrotron spectrum shapes.
Number spectrum in solid black and power spectrum in dashed red.
}
   \label {sr spectrum}
 \end{center} 
\end{figure*}

We can finally derive in the large radius ($\rho$) limit 
a compact result for the energy spectrum and the total rate:
\begin{eqnarray}
&&
\frac{d\Gamma}{d\omega} = 
N \sqrt{\frac{2\pi}{3}} \frac{Q^2 \alpha }{ 4\pi} 
\frac{1 }{ \gamma^2}
\int_{\omega/\omega_c}^{\eta\sqrt{\eta}\omega/\omega_c } d\xi
\frac{e^{-\xi}}{\sqrt{\xi}}
\left( (\frac{\omega_c \xi }{\omega })^{2/3} -1
\right) 
\,, \hspace{0.5cm}
\omega_c = \frac{3}{2\rho}\gamma^3 \sim 75 {\rm eV}  (\frac{\gamma}{10^4})^3 \frac{4 {\rm km}}{\rho}
\,,
\\ &&
\Gamma =  \sqrt{\frac{3}{2}}\frac{Q^2 \alpha }{8} \frac{1}{\rho}N\gamma
\sim 8.4 \times 10^{24} {\rm Hz} Q^2 \frac{4 {\rm km}}{\rho}
\frac{N}{10^{19}}\frac{\gamma}{10^4}
\,,
\end{eqnarray}
where $Q e$ is the charge of ion.
The value of the total rate given here corresponds to 1C ions equivalent to $\sim 10^{19}$ ion numbers.
The value $\eta$ in the upper bound of $\xi$ integral is estimated around $5$ from the available angular area of $4\pi$.
This result is in a fair agreement with the standard results
given in textbooks such as \cite{jackson},
considering the crudeness of matrix element estimate given here. 
We show the spectrum in Fig(\ref {sr spectrum})
for the reader's reference.

It is interpreted that 
for electron synchrotron radiation
the Zeeman splitting energy $eB/(\gamma m_e) = 1/\rho$ is extended by the $\gamma^3$ factor.
Use of the internal atomic energy in our problem has two important effects:
(1) larger energy spacing than the Zeeman splitting,
and (2) kinematical power law cutoff at $2\epsilon_{eg} \gamma$ rather than the exponential cutoff
$eB \gamma^2/m_e = \gamma^3/\rho \sim 2 \gamma^3{\rm neV} (100 {\rm m}/ {\rho} )$ 
for synchrotron radiation.

One may work out a requirement on the coherence $\rho_{eg}(0)$
by demanding that the synchrotron radiation is not an obstacle against the neutrino pair
emission  \cite{yokoya}.
It is  imposed that the number of emitted neutrino pair per revolution of
circular motion $\Gamma \times 2\pi \rho/c$ 
(equivalent to the number of de-excited ion caused by
neutrino pair emission) is much larger than the number of emitted synchrotron photon
per revolution.
This condition gives a constraint on the coherence,
\begin{eqnarray}
&&
|\rho_{eg}(0) | \gg 1 \times 10^{-4} Q (\frac{\gamma }{ 10^4})^{-3/2} (\frac{\rho }{4{\rm km} })^{-3/4}
 (\frac{ \epsilon_{eg} }{ 50 {\rm keV}})^{-11/4}
\,,
\end{eqnarray}
taking the spin factor to be unity, $S_e^2 = 1$.
If this condition is violated in the case of 
a large $Q$, one may
have to think of  compensating the loss of excited ions by
irradiation of laser each time of revolution \cite{excitation of atoms}.

We have several comments based on  results of the gamma emission
from excited ions.
First,  a different kind of coherence effect over a larger volume
may further enhance photon emission rates by
the super-radiance mechanism of Dicke \cite{sr}.
In the case of two-photon emission mentioned above
the macro-coherent paired super-radiance (PSR) may
further enlarge the coherent region \cite{renp overview}
not restricted to an area of the photon wavelength in the Dicke case.
It might even be possible to produce coherent
gamma ray 'laser', with a help of macro-coherence.
As an example, $2s \rightarrow 1s$ two-photon transition
of H$^-$ ion 
may be an excellent source of coherent two-photon emission
due to its long lifetime of $2s $ excited ion.
The achievable energy is not large, however, of order 200 keV $\gamma/10^4$
for hydrogen ion.
Molecular vibrational transition $v=1 \rightarrow 0$
of pH$_2^+$ may be better due to their easiness of
Raman excitation.
Recently, the macro-coherent PSR of vibrational 
transition of neutral pH$_2$ was observed \cite{psr observation},
in which we achieved a macro-coherence of $\sim $ several
\% over a target of 15 cm long.
Rates of  two-photon emission from circulating excited ions may be worked out
as in the rate calculation of neutrino pair emission.
Our $\gamma-$scaling law suggests that two-photon emission 
rates are large despite their effective, weaker coupling.

Even as a technical strategy towards high intensity neutrino beam, it would be wise to first study
basic experimental feasibility of
heavy ion excitation aiming at high energy photon emission,
since it would be easier to detect and study the mechanism of photon
emission in detail.

\vspace{0.5cm}
In summary,
a new method of producing CP-even coherent neutrino beam from
circulating excited ions was proposed.
Large production rates of neutrino energies extending to
much beyond the keV region were derived.
When ions are excited to a different E1 allowed level,
they may provide high intensity gamma ray beam
much beyond the keV range.
Evidently, much R and D works, both
theoretical and experimental, are needed to
determine a realistic design using a specific ion.

\vspace{0.5cm}
 {\bf Acknowledgements}

We should like to thank 
M. Yoshida, H. Sugawara, and the members of
Okayama SPAN group for delightful discussions.


\begin{thebibliography}{99}
\bibitem{excitation of atoms}
An interesting possibility of ion excitation is
to inject ions in the ground state into the ring,
and then to irradiate (for excitation) laser from
the counter-propagating direction  in the ring.
This way one may effectively boost the  laser energy 
 into the keV region, since ions see photons boosted
by the $\gamma$ factor.
Continuous irradiation of laser may also help
to maintain a high coherence of large $\rho_{eg}$.
\bibitem{uetake}
We appreciate S. Uetake for giving us valuable 
information on M1 decay rates of He-like ions.
\bibitem{yokoya}
We should like to thank K. Yokoya for pointing out
the possibility of a large beam loss due to the
synchrotron radiation for a large heavy ion charge.
\end{thebibliography}
\end{document}